\documentclass[letterpaper]{article} 
\usepackage[draft]{aaai2026}  
\usepackage{times}  
\usepackage{helvet}  
\usepackage{courier}  
\usepackage[hyphens]{url}  
\usepackage{graphicx} 
\usepackage{natbib}  
\usepackage{caption} 
\usepackage{algorithm}
\usepackage{algorithmic}

\usepackage{newfloat}
\usepackage{listings}
\DeclareCaptionStyle{ruled}{labelfont=normalfont,labelsep=colon,strut=off} 
\floatstyle{ruled}
\newfloat{listing}{tb}{lst}{}
\floatname{listing}{Listing}
\usepackage{amsmath}
\usepackage{booktabs}
\usepackage{xcolor}

\title{Assessing the Quality of AI-Generated Exams: A Large-Scale Field Study}
\author {
    Calvin Isley\textsuperscript{\rm 1},
    Joshua Gilbert\textsuperscript{\rm 2},
    Evangelos Kassos\textsuperscript{\rm 1},
    Michaela Kocher\textsuperscript{\rm 1},
    Allen Nie\textsuperscript{\rm 3},
    Emma Brunskill\textsuperscript{\rm 3},
    Ben Domingue\textsuperscript{\rm 4},
    Jake Hofman\textsuperscript{\rm 5},
    Joscha Legewie\textsuperscript{\rm 6},
    Teddy Svoronos\textsuperscript{\rm 1},
    Charlotte Tuminelli\textsuperscript{\rm 1},
    Sharad Goel\textsuperscript{\rm 1}
}
\affiliations {
    \textsuperscript{\rm 1}Harvard Kennedy School, Harvard University, Cambridge, MA 02138, USA\\
    \textsuperscript{\rm 2}Harvard Graduate School of Education, Harvard University, Cambridge, MA 02138, USA\\
    \textsuperscript{\rm 3}Computer Science, Stanford University, Stanford, CA 94305, USA\\
    \textsuperscript{\rm 4}Stanford Graduate School of Education, Stanford University, Stanford, CA 94305, USA\\
    \textsuperscript{\rm 5}Microsoft Research, Microsoft, New York City, NY, 10012\\
    \textsuperscript{\rm 6}Sociology, Harvard University, Cambridge, MA, 02138\\
    cisley@g.harvard.edu, jlegewie@fas.harvard.edu,
    ekassos@hks.harvard.edu, 
    mkocher@hks.harvard.edu, \\
    anie@cs.stanford.edu, 
    ebrun@cs.stanford.edu,
    bdomingu@stanford.edu, 
    jmh@microsoft.com, \\
    jlegewie@fas.harvard.edu, theodore\_svoronos@hks.harvard.edu,
    charlotte\_tuminelli@hks.harvard.edu,
    sgoel@hks.harvard.edu
}

\begin{document}

\maketitle

\begin{abstract}
While large language models (LLMs) challenge conventional methods of teaching and learning, they present an exciting opportunity to improve efficiency and scale high-quality instruction. One promising application is the generation of customized exams, tailored to specific course content. There has been significant recent excitement on automatically generating questions using artificial intelligence, but also comparatively little work evaluating the psychometric quality of these items in real-world educational settings. Filling this gap is an important step toward understanding generative AI's role in effective test design. In this study, we introduce and evaluate an iterative refinement strategy for question generation, repeatedly producing, assessing, and improving questions through cycles of LLM-generated critique and revision. We evaluate the quality of these AI-generated questions in a large-scale field study involving 91 classes---covering computer science, mathematics, chemistry, and more---in dozens of colleges across the United States, comprising nearly 1700 students. Our analysis, based on item response theory (IRT), suggests that for students in our sample the AI-generated questions performed comparably to expert-created questions designed for standardized exams. Our results illustrate the power of AI to make high-quality assessments more readily available, benefiting both teachers and students.
\end{abstract}

\begin{links}
    \link{Code}{https://github.com/calisley/ai_exams}
    \link{Datasets}{https://github.com/calisley/ai_exams/tree/main/data}
 \end{links}

\section{Introduction}
Generative AI presents both unique challenges and opportunities for educators. 
On one hand, it threatens to undermine traditional forms of assessment, if used improperly \cite{ho_artificial_2024,briggs_strive_2024}.
Moreover, potentially biased or otherwise inaccurate information produced by LLMs
could be especially harmful in educational settings, where students are particularly receptive to internalizing the provided information---even if it is wrong.
Despite these challenges, generative AI has the potential to dramatically improve the quality and accessibility of education. 
LLMs can in principle deliver highly personalized educational experiences at scale, democratizing access to high-quality instruction. 

Here, we investigate the potential of AI to assist with one specific aspect of teaching and learning: producing high-quality assessments customized to individual courses.
Creating suitable exam questions can take instructors and other experts hours or even days \cite{attali_interactive_2022,drori_human_2023}. For resourced-constrained educators, this time-consuming process may detract from their ability to meaningfully engage with students in other ways. The  alternative is for educators to spend little time on test creation, but this can result in tests that are too easy, too hard, or generally do not adequately capture student ability and aggregate class performance.

To address this issue, educational researchers have long explored ways to generate high-quality assessments automatically \cite{irvine_item_2002, gierl_automatic_2016}. The advent of LLMs represents a new frontier in automated question generation \cite{tan2024review}, with a flurry of recent activity seeking to address the problem.
However, despite much work exploring the potential of generative AI to create exam questions, there are relatively few studies formally studying the performance of LLM-generated items in real-world educational environments---a critical step for understanding and responsibly using these AI-generated questions.

In this study, we introduce an iterative refinement strategy---similar to Self-Refine \cite{madaan_self-refine_2023}---for generating exam questions tailored to specific course content. 
In one of the largest field studies of AI-generated assessments to date, we evaluate the quality of the generated exam items in 71 of our 91 participating college courses, reaching approximately 1,200 students. We use the remaining 20 courses (approximately 500 students) to benchmark the performance of our AI-generated questions against problems created by human experts for a standardized Advanced Placement (AP) exam. Full details of the field study are provided below. 

Using instructor-provided course materials, we tailor the AI-generated questions to each course's specific context. 
Analyzing the student responses with item response theory \citep[IRT;][]{van2016handbook}, 
we find that the AI-generated questions performed on par with those created by experts, both in terms of their overall difficulty and their discriminative power.
On average, the AI-generated items we produced were somewhat easier but also more discriminating than the expert-produced questions.
Our findings illustrate the potential for generative AI to increase access to high-quality, customized assessments, benefiting both teachers and students. 




\section{Related Work}
\subsubsection{Automated Item Generation } Since the advent of computer-based assessment, educational researchers have explored the potential of Automated Item Generation (AIG) \cite{irvine_item_2002, gierl_automatic_2016}. Most work on AIG has focused on college-level multiple-choice questions in STEM fields, a tradition we follow in this study \cite{song_automatic_nodate}. Historical approaches to AIG were often ``model" based, utilizing human-vetted schemas of questions containing variables that are later populated algorithmically~\cite{bejar_generative_2002, gierl_using_2012,attali_automatic_2018}. As a simple example, a schema might be of the form $a + b=?$, where the variables $a$ and $b$ could be assigned values.

\subsubsection{AI-Assisted Automated Item Generation}
Beyond this model-based approach, researchers have used natural language processing to extend AIG to domains and question-types that may be trickier to model, like factoid questions or personality test items \cite{von_davier_automated_2018, serban_generating_2016}
Following the recent development of LLMs, researchers have now shifted their attention towards using generative AI to produce items~\citep[e.g.,][]{attali_interactive_2022,russell-lasalandra_generative_2024,chan_automatic_2025,bhushan2025can}.
This work has
highlighted the value of proper prompt engineering for creating high-quality questions \cite{kiyak_chatgpt_2024}, but has not typically considered more advanced generation techniques, like the iterative refinement approach we apply here.
Past work has also left unanswered the extent to which generative AI can create highly tailored questions across a diverse range of course content.

Despite increasing interest in using generative AI to automatically generate items, 
there has been limited evaluation of item quality.
Most such work has relied exclusively on human experts to evaluate items, rather than considering actual responses to generated questions administered in field settings. Indeed, a recent review of 60 papers that used LLMs to perform AIG \cite{tan2024review} noted that psychometric evaluations of test items is not done in most papers, and ``recommend[ed] evaluating both the measurement properties and pedagogical
soundness of generated items as an essential step in AIG." There are, however, some exceptions to this general trend. 
Prior work~\cite{zelikman2023generating} used LLMs as simulated students to help automatically create parallel tests using a much smaller set of human data, with similar psychometric properties to human experts. Similarly, \citet{liu2025leveraging} recently examined using LLMs as simulated humans to help designers generate items, and evaluated whether the LLM responses yielded similar psychometric item parameters to those obtained using human responses. 
Additionally, two recent studies had humans take tests with both human and LLM-generated items, finding that both sets of items had similar item difficulty distributions in an IRT analysis.
While informative, these studies were done at a relatively small scale. 
The first was a 36 student participant study on English Language Teaching, finding that ChatGPT-generated items had slightly lower discrimination~\cite{kiyak_chatgpt_2024}; the second was a 207-subject study (outside the classroom) on Algebra, finding LLM-generated items had slightly higher discrimination~\cite{bhandari2024evaluating}. To our knowledge, there has not been a large-scale field study of the psychometric properties of generative-AI created items across a diverse range of college-level content. 


\section{Methodology}
Building on past work on AI-assisted AIG, we developed a two-stage, iterative process for generating customized, high-quality exam questions tailored to a class's specific course content.
Our goal was to create relatively short exams, comprised of 10 multiple-choice questions that could be completed in about 1-2 minutes each, to ease burdens on student participants.
Below we describe our AI-based exam generation process, as well the creation of our benchmark tests. 

\subsection{AI-based exam generation}
\begin{figure*}
    \centering
    \includegraphics[width =0.75\linewidth]{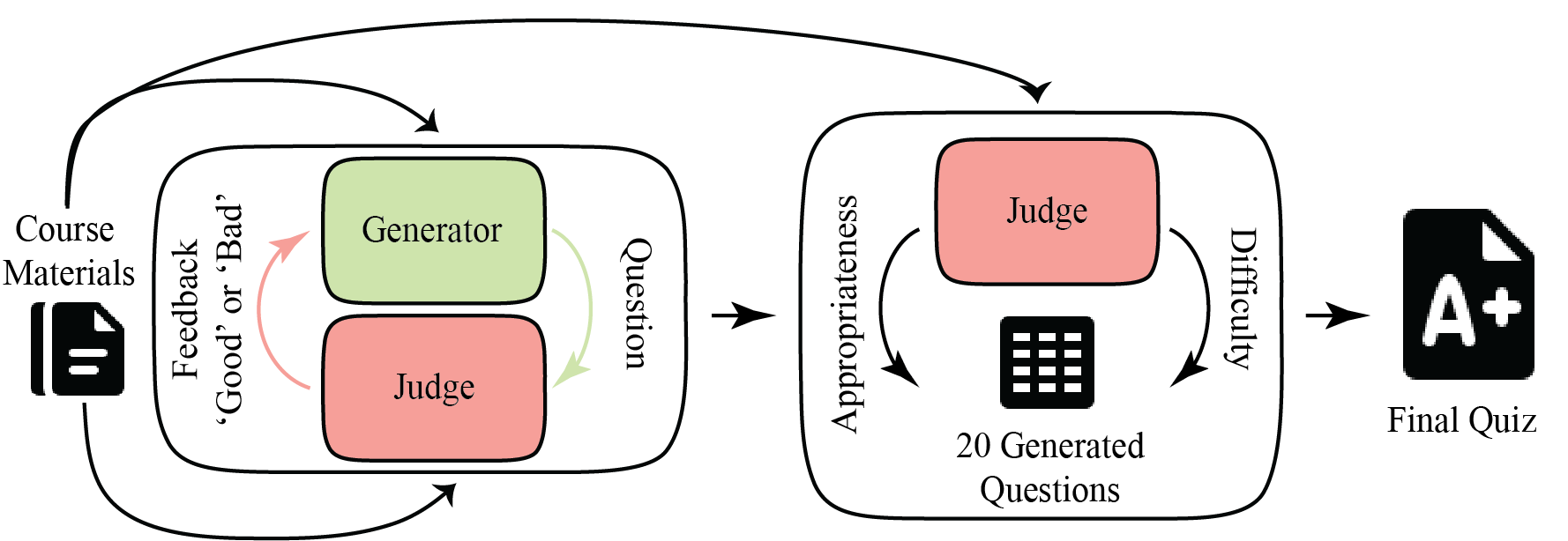}
    \caption{To produce our test items, our algorithm repeatedly generates and evaluates model outputs, until questions are deemed acceptable. Course materials are provided to both the AI-generator and AI-judge to ensure questions and feedback are context specific. After 20 questions are generated, they are passed to an additional AI-judge, who evaluates the appropriateness and difficulty of the 20 generated items. The hardest 10 become the final test.}
    \label{fig:workflow}
\end{figure*}
To generate exam questions, we employed an iterative prompt refinement strategy, similar to Self-Refine \cite{madaan_self-refine_2023}. 
To ensure questions were tailored to the covered class material, we incorporated a range of instructor-provided course materials, as described below.
All question generation and judging requests were performed via OpenAI’s o3-mini model---OpenAI's most advanced reasoning model available at the time of generation. A broad overview of the exam generation procedure is shown in Figure~\ref{fig:workflow}, and psuedo-code is provided in the appendix.

\subsubsection{Question Generation} 
For each class, we independently generated questions as follows.
To start, we compiled the instructor-provided course materials---including the course description, syllabus, and prior homework or exams---and converted them to plain text. This material was then included as context in the prompts for both the question generator and the judge. Full prompt templates are included in the appendix. 

Next, we repeated the following three steps until we had 20 judge-labeled ``good'' questions:

\begin{enumerate}
    \item \textbf{Question Generation: } Generate one multiple choice question that tests the concepts covered in the provided materials. To align the AI-generator to our desired question quality, we provided good and bad examples for few shot inference. 
    Specifically, in every iteration for all classes, our prompt contained five questions selected from the public 2012 AP Statistics practice exam as ``good" candidate examples.\footnote{The 2012 AP Statistics Practice exam can be found here: https://apcentral.collegeboard.org/media/pdf/ap-statistics-practice-exam-2012.pdf} No ``bad" questions were provided at the beginning of iteration. As the iterative refinement cycle progresses, previously generated questions are inserted into the prompt for future generation. We insert up to the first five ``good" and five ``bad" questions into the prompt, with the AI-judge's label indicating a question's quality (as described in Steps 2 and 3 below).
    
    \item \textbf{Question evaluation: } Assess the question's appropriateness for the class, uniqueness relative to the ``good" examples provided, and a few other functional criteria (e.g., ensuring that questions do not test students on information included in the syllabus). The AI-judge returns either a ``good" or ``bad`` label of question quality.
    
    \item \textbf{Prompt self-refinement: } 
    The question and label generated in steps 1 and 2 above are inserted into the AI-generator's prompt as additional context (i.e,. the previously generated question is provided as either a ``good" or ``bad" example).
\end{enumerate}

The following are examples of a ``good'' and a ``bad'' question produced by our iterative generation procedure. The ``bad'' question was rejected because it tests knowledge of logistical details of the class rather than substantive content.

\begin{quote}
    \textbf{``Good" --} \textit{Programming Languages}: In a statically-typed functional language that is extended with object-oriented features, what is a primary design challenge that arises from such integration?
\end{quote}

 \begin{quote}
 \textbf{``Bad" -- } \textit{Advanced Biochemistry Laboratory}:
    According to the final report guidelines for Advanced Biochemistry Laboratory, which of the following mistakes would most likely result in your manuscript being returned ungraded?
\end{quote}

When the judge rates that 20 good questions are generated, the loop above terminates. 
\subsubsection{From questions to exams} To create the full 10-question, class-specific exam, the 20 questions generated above were passed through a final round of AI-judging, which assessed question difficulty, appropriateness, and confirmed the generator-provided answer was correct. Any questions deemed inappropriate for the course and any question where the judge disagrees about the correct answer were removed. Finally, we then selected the 10 hardest questions (as evaluated by the AI-judge) to include in the exam. This was done because in our informal assessments, the generated questions appeared to be on the easier side, and so we prioritized the most difficult questions for inclusion in the exam; see the evaluation section below for further details on question difficulty.
If the final round of judging removed more than 10 questions, an additional 20 questions were generated for that course and the full group of 40 was re-evaluated. 
In the appendix, we include the full 10-question exam for a class on programming languages. 

\subsection{Benchmarking against traditional exam questions}
We compared our AI-generated exam questions to high-quality human-generated questions designed for use on standardized tests. In particular, we benchmarked our method against the 24 multiple-choice questions included in a publicly available 2012 AP Statistics practice exam that could be faithfully recreated with plain text (i.e., did not include figures or tables). 

The range of topics tested by these questions, as well as their general difficulty, appeared to align well with the material covered in the 20 statistics courses in our study.
To further tailor the exams to the classes, 
we used an LLM to select the 10 most appropriate questions for each statistics class from this question bank. 
Specifically, we used o3-mini to first assess a question's key concepts and difficulty. For each of the statistics courses, we then constructed a list of concepts covered in the class, which we extracted from the provided class materials. Finally, questions were added to the exam in descending order of difficulty---on the theory that the college classes in our sample were more difficult than typical AP classes offered in high school---while preserving course fit and concept diversity. 

We limit our benchmark to statistics-related exams, as this was one of the only areas for which we could find high-quality traditional questions that tested the material covered by the classes in our study. While this limits the generalizability of our comparison, it also illustrates our key contribution: automatic creation of highly tailored questions across a diverse array of subjects for which there are often few alternatives.

\section{Evaluation}
We assessed the quality of our AI-generated exams, as well as the benchmark exams, via a large-scale field study conducted at a diverse range of colleges and universities across the United States. Student responses were analyzed with item response theory (IRT), yielding assessments of each question in our corpus, as well as the overall quality of the class-specific exams.

\subsection{Field Study}
Our field study comprised 182 classes at American colleges. These classes corresponded to 164 unique courses, as some courses enrolled multiple sections of the same course as unique classes. This sample primarily included courses in STEM related fields. Instructors provided their course materials to facilitate tailored question generation. Neither students nor instructors were told whether questions were AI-generated or came from a standardized test. 

At the beginning of the 2025 Spring semester, students in participating classes took a common pre-test of general quantitative reasoning skills. This common assessment was used to obtain comparable estimates of ability across our diverse student population.
Then, near the end of the semester, students completed the tailored exam that we had created for their class. 
Participating instructors had the choice to assign both the pre-test and semester-end exams as either homework or as an in-class assignment. All sections of the same course received the same semester-end assessment. 

Of the 182 participating classes, we have exam responses from 121 courses (66\%). For the remaining courses, we either had low student participation rates or determined that the course itself was not suitable for exam generation (e.g., it was a research seminar without homework or other exams).
Of the 121 classes for which we ultimately collected exam data, we manually created exams for 30 classes (24\%) through an ad-hoc process to pilot our exam-administration procedure; we excluded these classes without analyzing the corresponding data. 

Our final dataset thus consists of responses from 91 classes, comprising 1686 students who completed both the pre-test and semester-end exam.
In sum, 1208 students in 71 of these classes received the AI-generated exam, and the remaining 478 students in the 20 statistics classes received an assessments with questions from the AP statistics practice exam. 
In some rare cases (less than 2\%), students submitted multiple responses for either the pre-test or semester-end exams. In these instances, we superscored their responses, 
retaining their highest score for each individual question across attempts. 

\subsection{Item Response Theory}
We evaluate the quality of both individual exam questions and the complete 10-question exams with item response theory (IRT). 
IRT \cite{van2016handbook} is a standard method for obtaining assessments of question (referred to as an ``item'') and exam quality.
It yields estimates of both \emph{difficulty} (i.e., how hard an item is) and \emph{discriminatory power} (i.e., how well an item separates out stronger from weaker students).

IRT models the probability of a correct response as a function of an individual's latent ability $\theta_i$ and latent properties of each item. We use a two-parameter logistic model (2PL).
Specifically, let $Y_{i,j}$ indicate whether item $j$ was correctly answered by student $i$. We then model,
\begin{equation*}
\Pr(Y_{i,j}=1 \mid \theta_i, \alpha_j,\beta_j)=\operatorname{logit}^{-1}[\alpha_j(\theta_i-\beta_j)]
\end{equation*}
where $\alpha_j$ and $\beta_j$ are latent item parameters.

An item's difficulty parameter $\beta$ indicates the ability level $\theta$ at which $\Pr(Y =1 \mid \theta, \alpha, \beta)= 0.5$. As an item's difficulty increases, only students with higher latent ability parameters are expected to answer the item correctly. 
An item's discrimination parameter $\alpha$ captures how sharply $\Pr(Y=1 \mid \theta, \alpha, \beta)$ changes as a function of $\theta$. More discriminating items do a better job at distinguishing between students whose abilities lie near the item's difficulty.

To evaluate exams on the whole, we compute each test's \emph{information curve}, which shows how much information a test provides about students across the spectrum of ability levels. 
The information curve $I_j$ of an item is: 
\begin{equation*}I_j(\theta)=\alpha_j^2 P_j(\theta) (1-P_j(\theta)),
\end{equation*}
as defined in~\citet{lord_applications_1980},
where $P_j(\theta) = \Pr(Y_j=1 \mid \alpha_j, \beta_j,\theta)$. To find an assessment's total test information, we sum over its constituent questions:
\[
I(\theta) = \sum_j I_j(\theta).
\]
The information of an exam at a given $\theta$ tells us the precision with which the exam measures ability at that point.
Namely, from test information one can also compute a test's conditional \emph{reliability} 
\[
R(\theta) = \frac{I(\theta)}{1+I(\theta)},
\] 
which equals the proportion of total variance in estimated ability that is true signal rather than measurement error at that ability level \cite{nicewander_conditional_2018}. 

We examine both the individual item parameters as well as the aggregate test information to robustly understand how the AI-generated exams performed in the field, and how AI-generated questions compare to traditionally created questions appearing on standardized tests. 

\subsubsection{Model Inference}
We apply the above IRT model to simultaneously estimate parameters for all students and items in our dataset, including both pre-test and semester-end responses in the same model. 
We include pre-test responses as anchor items, allowing us to establish a common $\theta$ scale between students who took the AI generated exams and those who took the exams composed of questions from standardized tests. Though exam type was not randomly assigned, the common $\theta$ allows us to make descriptive comparisons across the two groups. 
We note that our choice to use a single parameter $\theta$ for each student implicitly assumes their latent ability is constant throughout the time period we consider. We further comment on this assumption in the discussion.

In our setting, we have hundreds of items distributed across dozens of classes. To efficiently estimate student and item parameters, we use a 
Bayesian hierarchal formulation of the 2PL IRT model. 
Hierarchical priors allow us to pool information across items, accounting for variation in both item discrimination and difficulty simultaneously \cite{burkner_bayesian_2021,konig_benefits_2024}. 

Specifically, for the discrimination parameter $\alpha_j$, we assume,
\begin{equation*}
    \log(\alpha_j) \sim N(\mu_\alpha, \sigma_\alpha^2),
\end{equation*}
where 
\begin{align*}
\mu_\alpha &\sim N(0,1)\\
\sigma_\alpha &\sim \mathrm{half}\text{-}\mathrm{Cauchy}(0,1).
\end{align*}
We follow \citet{gelman_prior_2006, polson_half-cauchy_2012} in setting weakly informative half-Cauchy priors for the standard deviation hyper-parameters. 
Similarly, 
for the difficulty parameters $\beta_j$, we assume,
\begin{equation*}
    \beta_j \sim N(\mu_\beta, \sigma_\beta^2),
\end{equation*}
where, as above, 
\begin{align*}
\mu_\beta &\sim N(0,1)\\
\sigma_\beta &\sim \mathrm{half}\text{-}\mathrm{Cauchy}(0,1).
\end{align*}
Finally, we follow standard practice and assume individual abilities are distributed according to a standard normal distribution~\cite{konig_benefits_2024}: \begin{equation*}
    \theta_i\sim N(0,1).
\end{equation*}
which is necessary for our model to be properly identified. 

\begin{figure}
    \centering
    \includegraphics[width=\linewidth]{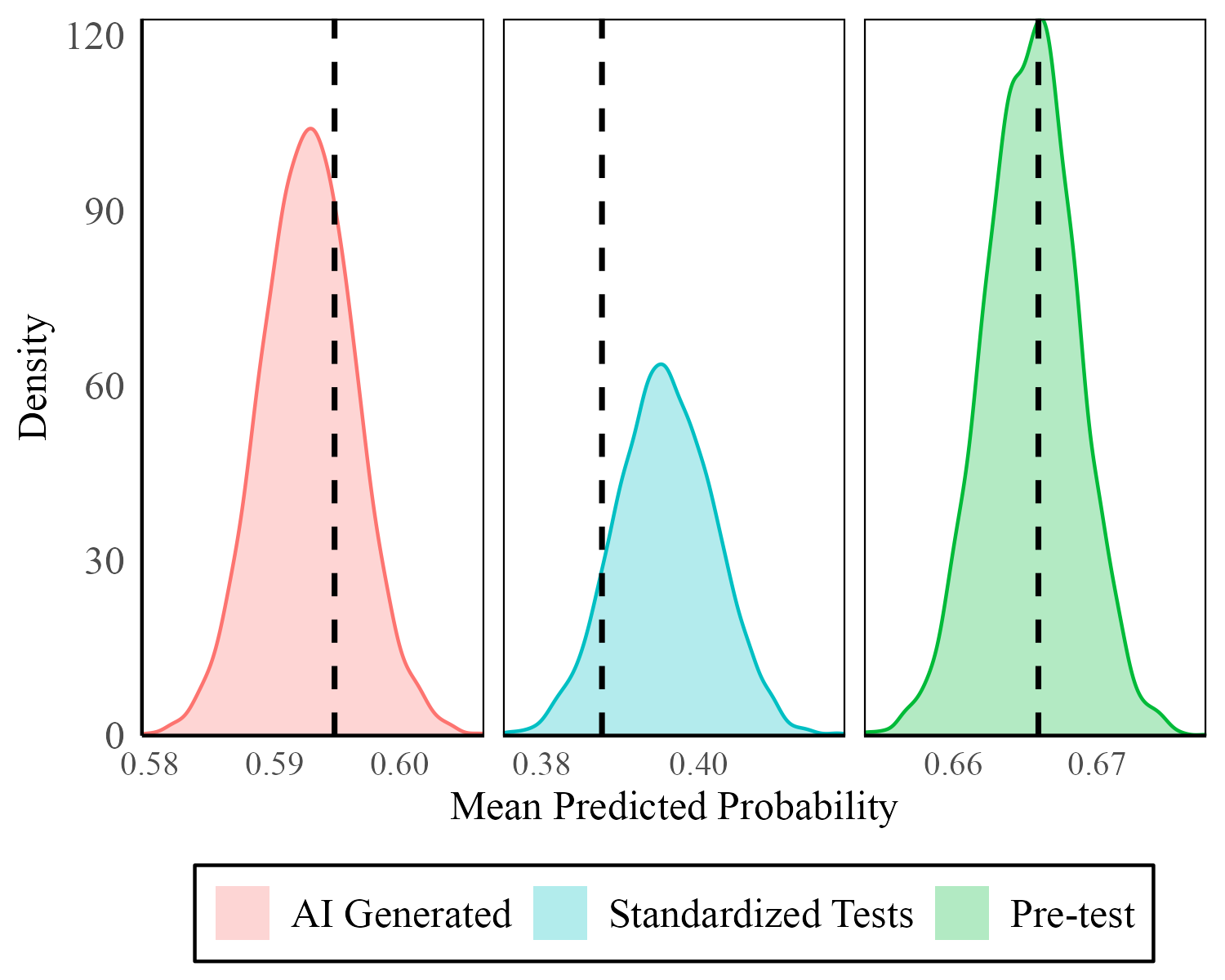}
    \caption{Distribution of predicted proportion of correct answers for questions of each type. The dashed line indicates the observed proportion of correct answers for participants in our field study for each exam type.}
    \label{fig:ppcs}
\end{figure}

We fit this hierarchical Bayesian IRT model using \texttt{cmdstanr} \cite{cmdstanr}, with 4 chains and 1000 samples per chain.
We confirmed that the Gelman-Rubin convergence diagnostic, $\hat{R}$ value of each parameter in the model was well below $1.05$ ($\max \hat{R} = 1.008$), indicating proper convergence ($>1.1$ indicates potential lack of convergence, $\approx 1$ indicates good convergence). 
All parameters achieved suitable effective sample sizes (ESS), with a minimum bulk ESS of 773 and a minimum tail ESS of 1,486-- indicating that our chains produced sufficient samples to estimate the posterior distributions reliably. Together these indicate that our chains mixed well and the posterior summaries are reliable.

We further conducted posterior predictive checks to investigate the calibration of our fitted model.
Specifically, we compared the posterior distribution of the proportion of questions answered correctly to the observed value.
The results are shown in Figure~\ref{fig:ppcs}, disaggregated by exam type: AI-generated, standardized test, and pre-test.
The observed proportion of correct answers from participants in our field study lie within the central mass of the posterior distributions, indicating the model is generally well-calibrated across exam types. 

As described above, we identify the model by assuming student ability follows a standard normal distribution: $\theta_j\sim N(0,1)$. 
Checking the posterior distribution of student abilities, we find it remains centered near zero with unit variance. This pattern indicates our items have induced neither excessive shrinkage nor undue dispersion in the ability estimates, suggesting the model is well suited to the available data. The full posterior distribution of estimated student abilities is available in the appendix. 

\subsection{Results}
We now assess the quality of our AI-generated and standardized exam questions---in terms of their difficulty, discrimination, and test information.

\subsubsection{Difficulty}
\begin{figure}
    \centering
    \includegraphics[width=\linewidth]{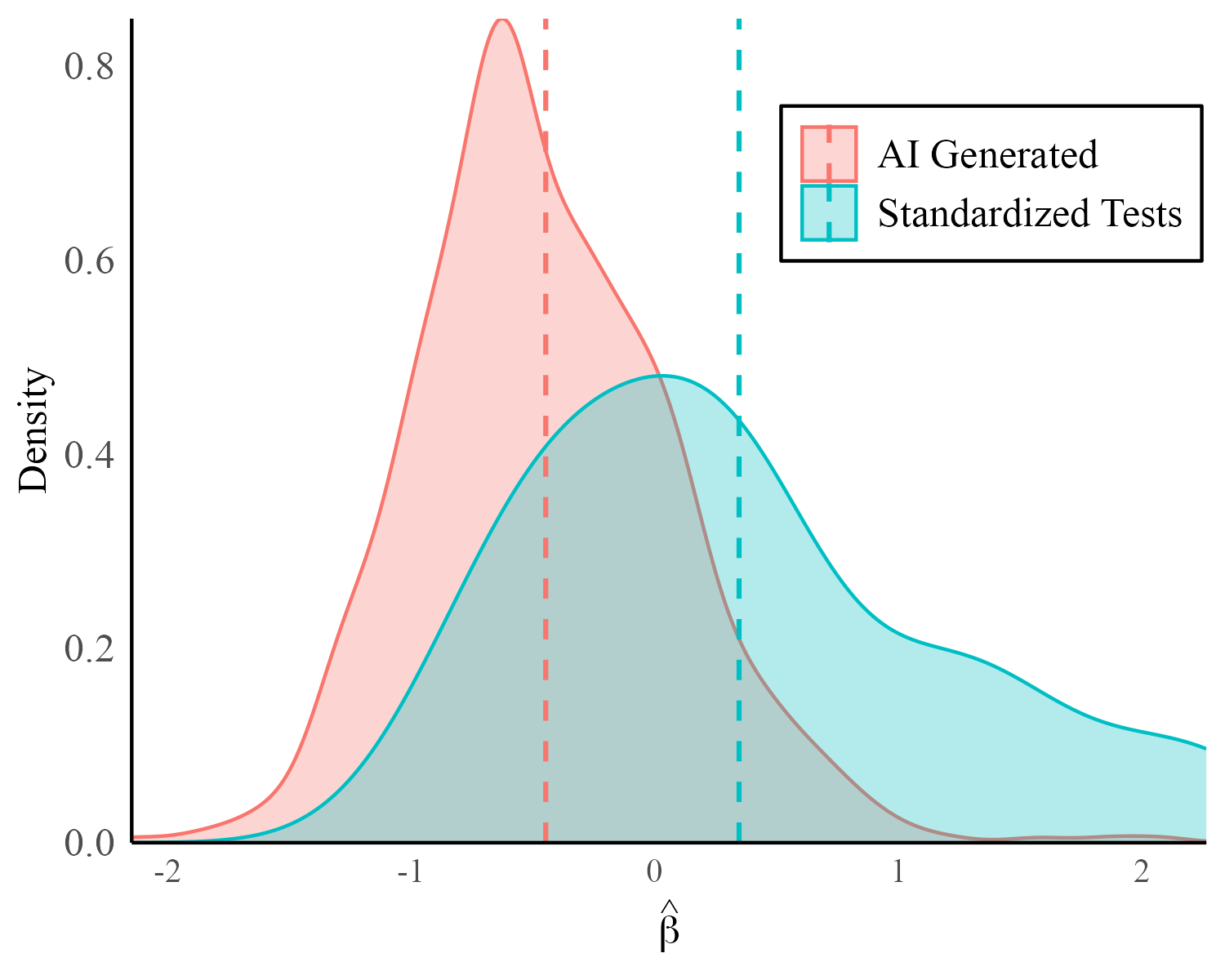}
    \caption{Distribution of the estimated difficulty parameters $\beta_j$ by type of exam, with dashed lines at their respective means.}
    \label{fig:difficulty}
\end{figure}

Figure~\ref{fig:difficulty} shows the distribution across items of the estimated difficulty parameters $\hat\beta_j$ for each question type. 
The average difficulty of the AI-generated questions is $\bar{\beta}_{\text{AI}} =-0.45$,
indicating that students with estimated abilities somewhat below the average had approximately even odds of answering these questions correctly.
This patterns aligns with the fact that
students correctly answered approximately 60\% of the AI-generated questions (Figure~\ref{fig:ppcs}), meaning that the typical student was expected to answer slightly more than half of these questions correctly.
This pattern may result from our question-generation process being anchored to instructor-provided exams and homework assignments, which might exhibit similar difficulty levels.

In contrast, the standardized test questions appear to be somewhat more difficult, with $\bar\beta_{\text{STD}} = 0.35$.
This number aligns with the lower overall correctness rate of these items compared to the AI-generated questions (39\% vs. 60\%, as shown in Figure 2).
The distribution of the average difference $\delta_\beta =\bar{\beta}_{\text{AI}}-\bar{\beta}_{\text{STD}}$ has posterior mean $E[\delta_\beta] = -0.79$, with 95\% credible interval $[-0.94,-0.65]$.  $\Pr(\delta_\beta<0)\approx 1$, indicating it is unlikely that the differences we observe are due to sampling variation. 
Our goal was to select questions that were appropriately calibrated to our student sample, though it is difficult to do so precisely in the absence of detailed student data.

\subsubsection{Discrimination}
\begin{figure}
    \centering
\includegraphics[width=\linewidth]{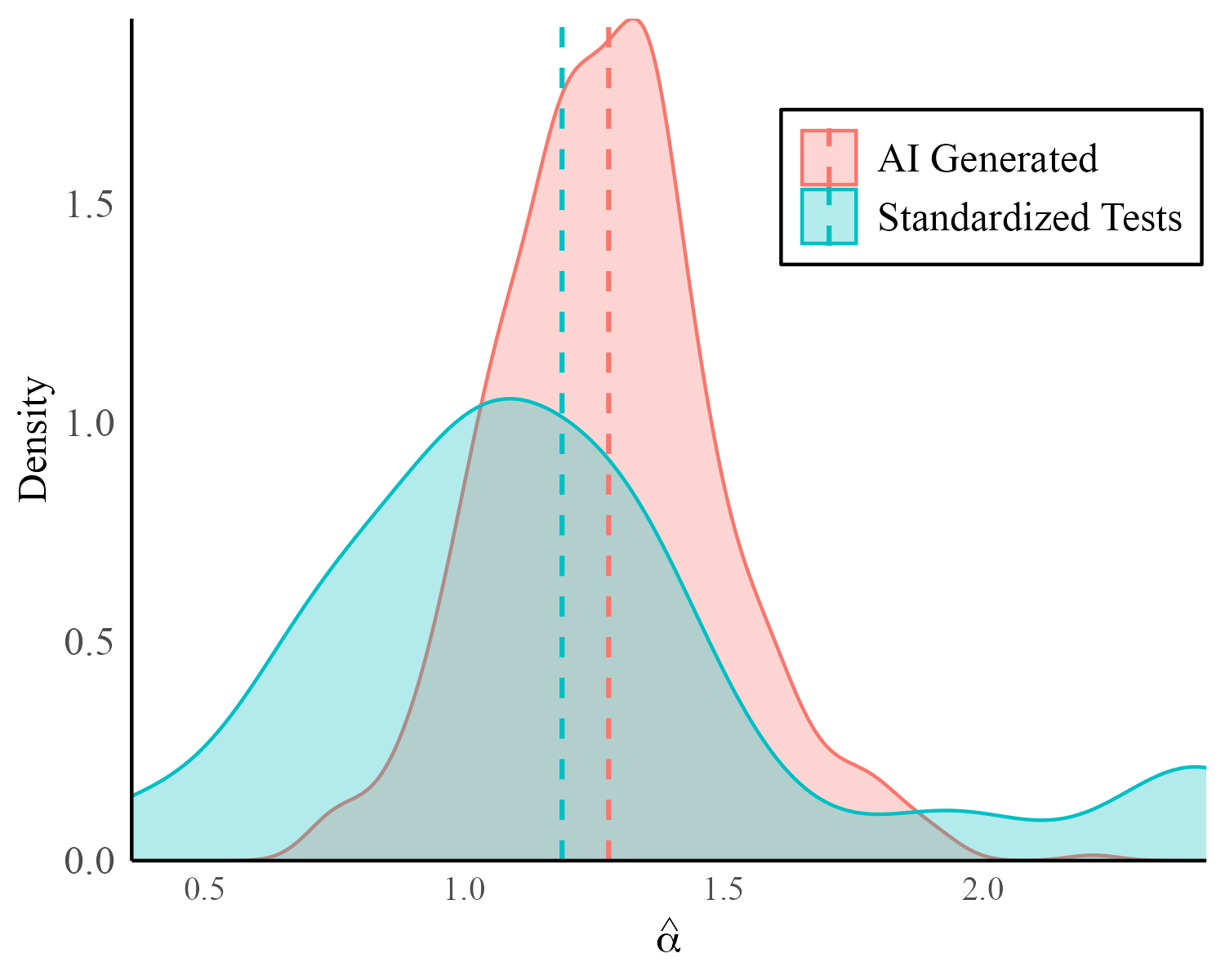}
    \caption{Distribution of estimated discrimination parameters $\hat{\alpha}_j$ across items, disaggregated by type, with dashed lines at their respective means.}
    \label{fig:discrimination}
\end{figure}
Analogous to our analysis above, 
Figure~\ref{fig:discrimination} shows the distribution of $\hat{\alpha}_j$ by item-type: AI-generated or standardized tests. 
The AI-generated questions had an average discrimination of $\bar\alpha_{\text{AI}} = 1.3$ across items. 
In contrast, standardized test questions were slightly less discriminating, with an average of $\bar\alpha_{\text{STD}} = 1.2$. 
By standard IRT thresholds \cite{baker_basics_2001}, both AI and standardized test questions yielded, on average, ``moderate" discriminating power.
The posterior distribution of their difference $\delta_\alpha =\bar{\alpha}_{\text{AI}}-\bar{\alpha}_{\text{STD}}$ has posterior mean $E[\delta_\alpha] = 0.09$, with $95\%$ credible interval $[-.0918,.253]$. Here, $\Pr(\delta_\alpha >0) \approx 0.85$ indicating moderate evidence that AI-generated questions are slightly more discriminating than standardized test items in our sample. 

\begin{table}[t]
    \centering
    \begin{tabular}{lrr}
        \toprule
        Discrimination & AI (\%) & Std. Tests (\%)\\
        \midrule
        Low: (0.35-0.64] & 0.0 & 8.3\\
        Moderate: (0.64 - 1.34] & 64.1 & 70.8\\
        High: (1.34-1.69] & 31.7 & 8.3\\
        Very high: ($>$1.7)& 4.2 & 12.5\\
        \bottomrule
    \end{tabular}
    \caption{Proportions of questions by discrimination level, following the taxonomy of \citet{baker_basics_2001}.}
    \label{tab:discrim_bins}
\end{table}

Table~\ref{tab:discrim_bins} shows the proportion of items at each discrimination level, following the taxonomy of \citet{baker_basics_2001}.
While a large share of both the AI-generated (64\%) and standardized test questions (71\%) are considered ``moderate", 36\% of the AI-generated questions are ``highly" or ``very highly'' discriminating compared to 21\% of the standardized test questions.
On the whole, the AI-generated questions are slightly more discriminating and more tightly distributed than the standardized test questions. We interpret these results as evidence that generative AI tools---and our iterative refinement procedure in particular---are capable of consistently constructing questions that can reliably differentiate between students. This suggests that instructors and testing platforms could leverage AI-driven item generation to create items with comparable discriminatory power at a fraction of the cost, and could offload creation of easier questions to an AI. 
\subsubsection{Test Information and Reliability}
We conclude by evaluating the quality of the aggregate 10-question exams for each class.
To do so, we consider each exam's 
test information curve and, in particular, the 
maximum information value associated with each exam.

\begin{figure*}[t]
    \centering
    \includegraphics[width=0.8\linewidth]{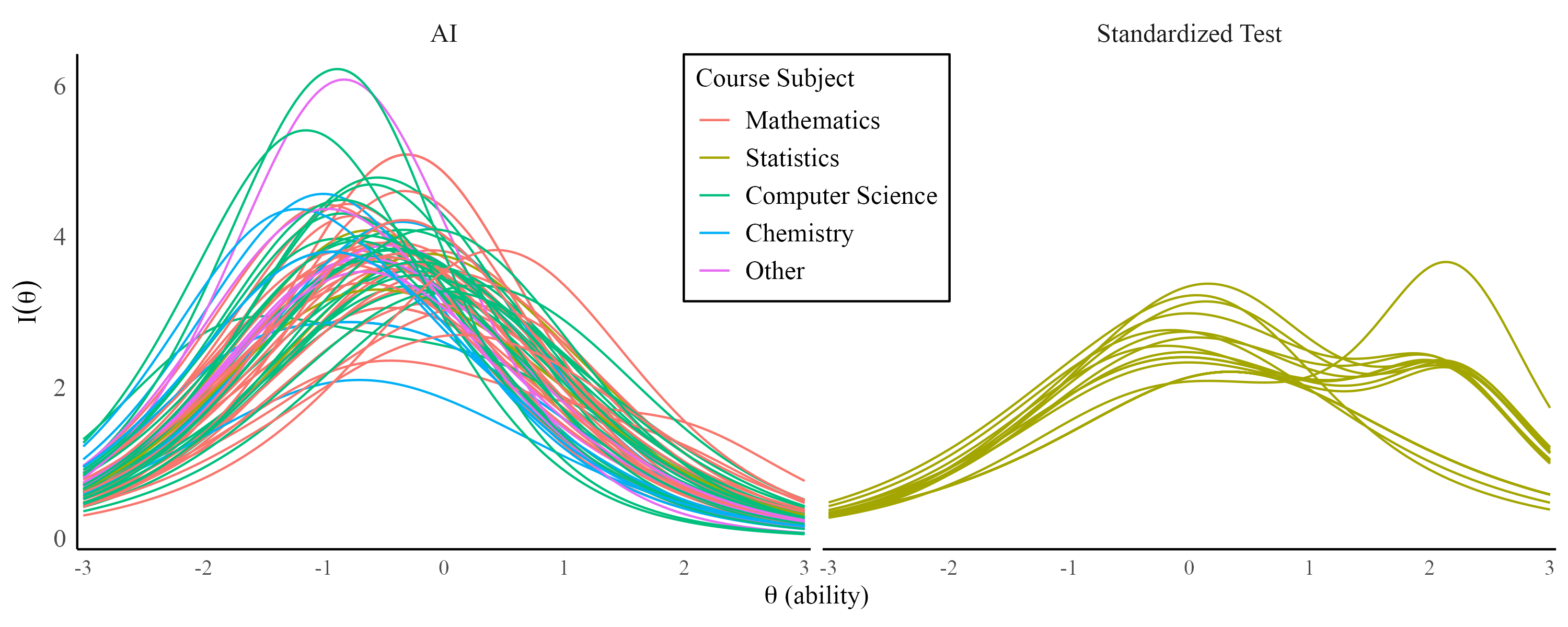}
    \caption{The test information curves for every class in our sample, split out by which type of exam each class received.}
    \label{fig:test_information}
\end{figure*}

Figure~\ref{fig:test_information} shows the test information curves for every class by exam type, colored by subject matter.
The maximum test information for the AI-generated exams, averaged over exams, is $I_{\max} = 3.85$, corresponding to a reliability of $R(\theta) = 0.79$. 
The test information curves for the AI-generated exams peaked at an average ability level of $\theta=-0.51$, indicating they were most sensitive for students with slightly below-average ability---which aligns with our finding that the AI-generated exams were best targeted to students of below average ability. 

In comparison, the standardized tests have lower max information, $I_{\max}=2.61$, with corresponding reliability $R(\theta) = 0.72$---in line with our results on the relative discrimination of AI-generated versus standardized test items. 
Further, these standardized exams peaked at an average ability level of $\theta=0.32$, suggesting they are most discriminating for students of slightly above average ability, again as expected from our difficulty results. 
We note that some of the standardized tests have bimodal information curves, likely due to these exams having a subset of particularly difficult questions (as shown in Figure~\ref{fig:test_information}).

Overall, the test information curves further suggest that the AI-generated exams are more discriminating than the standardized tests, while being maximally informative for students with somewhat lower inferred ability.
We note, however, that even though the AI exams show peak reliability for students of below-average ability, they still provide strong coverage for students across the ability spectrum. In particular, at $\theta = 0$, the AI exams have an average $I(0) = 3.36$, whereas the standardized test questions have an average $I(0)=2.5$. Indeed, for all $\theta<.75$, the AI generated exams were, on average, more maximally informative than the tests composed of questions from standardized exams for the students in our population, illustrating the strength of the AI generation. 

\section{Discussion}
Building on past work on automated item generation, we apply an iterative refinement strategy to create questions tailored to courses across a diverse range of subjects.
In one of the largest field studies of AI-generated questions to date, 
we found that this approach achieves psychometric properties on par with those of questions appearing on high-stakes standardized tests.
These results demonstrate the potential of AI to produce high-quality assessments at scale across varied topical domains. 

Our study is subject to some key limitations. First, we evaluated the standardized test questions exclusively for the statistics courses in our sample, reducing the generalizability of our comparison between AI-generated and standardized test questions. 
We note, however, that this limitation stems from the relative dearth of publicly available question banks that are suitable for idiosyncratic courses---illustrating the value of AI generation.
Second, for ease of grading, we limited our study to multiple-choice questions. 
Our question-generation procedure can, however, be straightforwardly modified to produce open-response or short-answer questions. 
Third, we limited our evaluation to relatively short, 10-question tests to avoid overburdening participants, though it seems likely that our results would extend to longer assessments.
Fourth, our reference condition approximated but likely did not fully replicate a human expert-written exam tailored to each statistics course. Specifically, although we used human-authored AP statistics questions, we still had to match these questions to each course based on concepts and difficulty as judged by an LLM, which may have introduced subtle differences from what a human expert might have done.
Finally, our assumption of a constant $\theta$ at two time points was necessary to establish of a common $\theta$ scale; inferences based on this assumption would be especially likely to be invalid if students in classes that received AI tests had larger (or smaller) changes in $\theta$ during the class relative to those in classes which took the human-generated exams.

Despite these limitations, our approach to exam generation appears promising for broader educational applications.
Future work could evaluate expert-generated questions in subjects beyond statistics, potentially drawing on materials that instructors themselves have created in the past.
For the most accurate inferences, one would ideally randomize assessments to be either AI- or human-generated. 
Moreover, one could expand beyond multiple-choice questions into short answer or open response, using an LLM to grade the open-ended responses.
Finally, to generate questions with specific difficulty and discrimination levels, one could  
fine-tune a model using the type of student response data that we've collected, improving 
question calibration.

More broadly, AI-assisted question generation has the potential to substantially reduce instructor workloads, increase access to high-quality assessments, and improve learning outcomes through more tailored instruction.
Instructors often have limited time and resources to prepare course materials; with AI-assisted exam generation, instructors may be able to spend more time on direct student instruction.
Further, our exam generation procedure can be immediately adapted to non-English speaking populations, and tailored to diverse cultural contexts, enhancing educational access. Finally, generating questions in real-time could facilitate tailored, adaptive practice, enhancing student engagement and learning outcomes.
Looking forward, we hope our work will contribute to more impactful, dynamic, and inclusive teaching and learning environments.
\section{Acknowledgments}
We thank the students who participated in our study and the instructors who facilitated the administration of our exams. Additionally, we thank WestEd for their assistance recruiting participants.
This work was funded by Microsoft 
and by the William F. Milton Fund at Harvard University.

\bibliography{draft_ai_exams}

\clearpage

\clearpage
\onecolumn
\section{Appendix}
\subsection{Gelman-Rubin convergence diagnostic Check}
\begin{figure}[h]
    \centering\includegraphics[width=.5\linewidth]{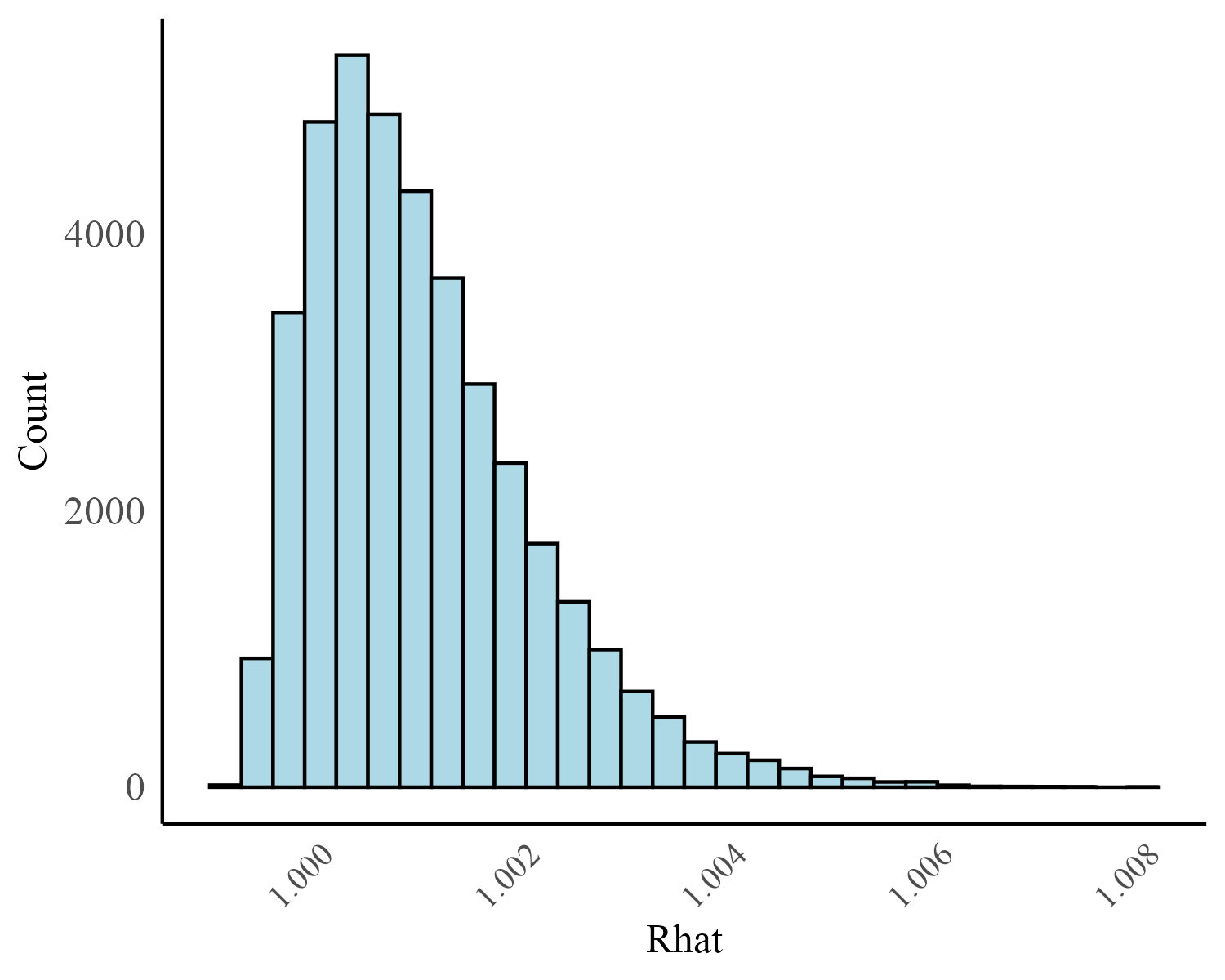}
    \caption{Gelman-Rubin convergence statistic for all parameters in our 2PL IRT model. }
    \label{fig:rhat}
\end{figure}
\subsection{Student Ability}
\begin{figure}[h]
    \centering
    \includegraphics[width=.5\linewidth]{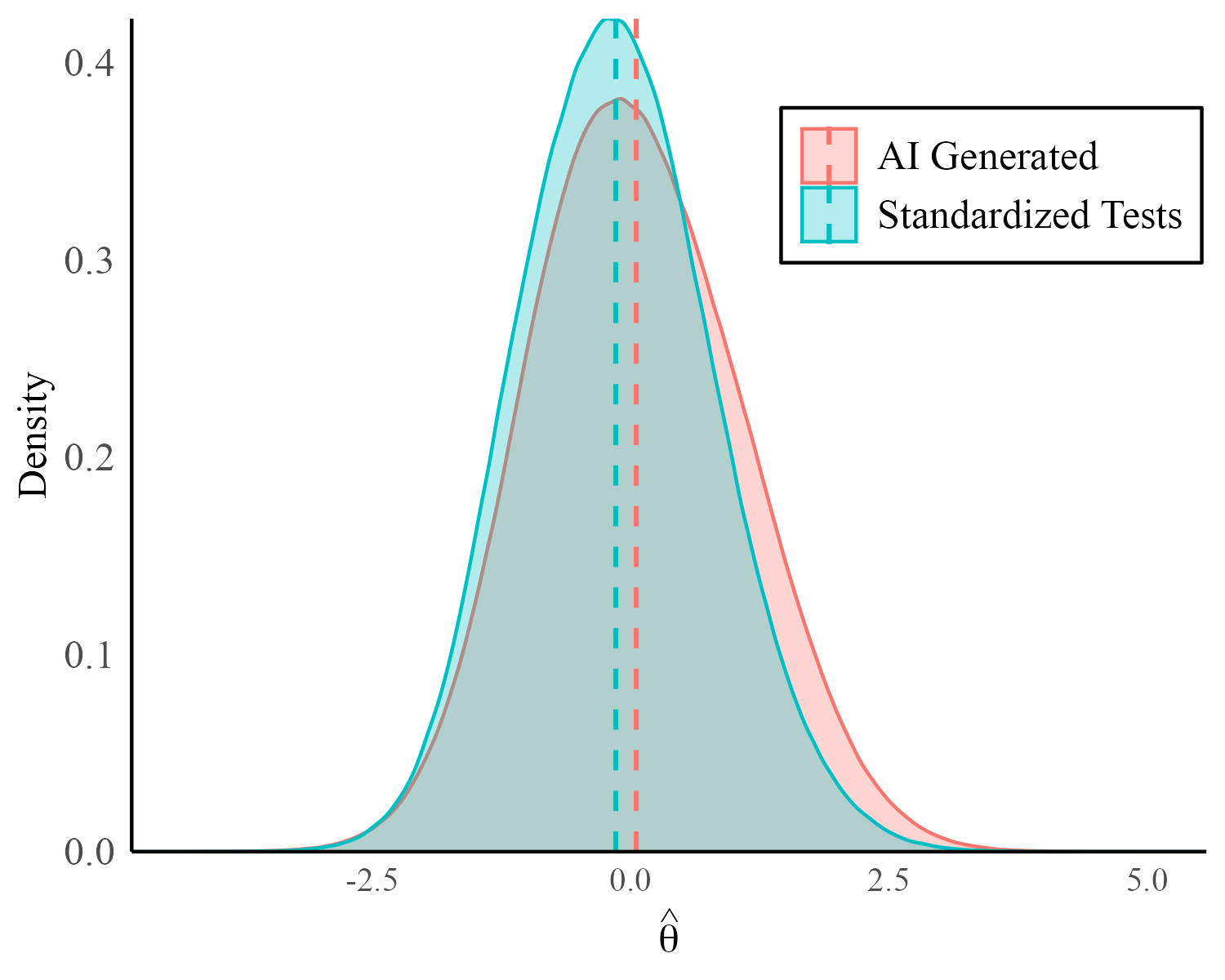}
    \caption{Distribution of the estimated student ability parameters $\theta_j$ by type of exam, with dashed lines at their respective means.}
    \label{fig:ability}
\end{figure}
\clearpage
\onecolumn
\subsection{Test Generation Pseudocode}

\begin{algorithm}
\caption{Assemble 10 High-Quality MCQs}
\begin{algorithmic}[1]
  \REQUIRE courseMaterials  \COMMENT{syllabus, description, past work}
  \ENSURE finalExam        \COMMENT{10 high-quality MCQs}

  \STATE \textbf{Pre-processing:}
  \STATE materialsText $\gets$ ConvertToPlainText(courseMaterials)
  \STATE generatorPrompt $\gets \langle\,\text{materialsText},\,\text{fiveGoodAPStatsExamples}\rangle$
  \STATE goodMCQs $\gets \emptyset$, \quad badMCQs $\gets \emptyset$
  
  \STATE
  \STATE \textbf{Generate–label–refine loop (until 20 good MCQs):}
  \WHILE{$|\text{goodMCQs}| < 20$}
    \STATE mcq $\gets$ GenerateMCQ(generatorPrompt)
    \STATE quality $\gets$ JudgeQuality(mcq, materialsText)
    \IF{quality = \texttt{"good"}}
      \STATE goodMCQs $\gets$ goodMCQs $\cup \{\,\text{mcq}\}$
    \ELSE
      \STATE badMCQs  $\gets$ badMCQs  $\cup \{\,\text{mcq}\}$
    \ENDIF
    \STATE generatorPrompt $\gets$ RefreshPrompt(\\
    \quad\quad generatorPrompt,\; first5(goodMCQs),\; first5(badMCQs)\\)
  \ENDWHILE

  \STATE
  \STATE \textbf{Exam assembly (difficulty filter \& top-up):}
  \STATE graded   $\gets$ JudgeDifficultyFitCorrectness(goodMCQs, materialsText)
  \STATE approved $\gets \{\,g\in \text{graded} \mid g.\text{isApproved}\}$
  \STATE ranked   $\gets$ SortDescendingByDifficulty(approved)
  
  \WHILE{$|\text{ranked}| < 10$}
    \STATE extraGood $\gets$ GenerateAdditionalGoodMCQs(\\
    \quad\quad generatorPrompt, materialsText, \text{count}=20\\)
    \STATE goodMCQs $\gets$ goodMCQs $\cup$ extraGood
    \STATE graded   $\gets$ JudgeDifficultyFitCorrectness(goodMCQs, materialsText)
    \STATE approved $\gets \{\,g\in \text{graded} \mid g.\text{isApproved}\}$
    \STATE ranked   $\gets$ SortDescendingByDifficulty(approved)
  \ENDWHILE

  \STATE finalExam $\gets$ \text{first 10 elements of ranked}
  \RETURN finalExam
\end{algorithmic}
\end{algorithm}
\subsection{Prompt Templates}

\begin{lstlisting}[
  basicstyle=\ttfamily,
  numbers=none,
  frame=lines, 
  xleftmargin=2pt,  
  breaklines=true,
  breakautoindent=false,
  breakindent=0pt,
  breakatwhitespace=true,
  title={Generator Prompt}
]
You have the following course information:

Course Name: {course_name}
Course Description: {desc}
Exam Content: {exam_content}
Syllabus Content: {syllabus_content}

Generate EXACTLY ONE multiple-choice question in JSON format, with these fields:
- question
- options (an array of exactly four answer strings)
- answer (the correct option text exactly)
- explanation
- relevant_courses
- difficulty (1-10)
- key_concepts (list of strings)
- question_type
- quality (1-10)

Guidelines:
- The question must be mentally solvable in <5 minutes.
- No direct references to external data or prior questions.
- Must be non-trivial: no purely formula-plugging or trivial difficulty.
- Return ONLY the JSON (no extra text).
- Do NOT generate a question that covers concepts already touched on in prior questions.
- Only use the provided materials as background for what concepts are covered in the course. Do not base your questions explicitly off of course materials. 

Importantly, do not ask questions about concepts that are not in the course materials. Do not infer concepts, only use what is presented in the materials. 

In general, question generation should follow the following guidelines:

1. Difficulty and Conceptual Depth:
- Multi-Step Reasoning: Questions should require several steps of reasoning or combine two or more concepts. For example, a question might ask students to analyze a scenario by connecting multiple theoretical ideas.
- Subtle Distractors: Create answer options that are closely related conceptually, with subtle differences, so that common misconceptions or partial understandings can lead to selecting an incorrect option.
- Conceptual Integration: Ensure that some questions merge multiple key topics or concepts from the course, challenging students to synthesize information rather than apply a single formula or idea.
- Non-Routine Scenarios: Include some problems with novel or less typical settings. This could involve re-framing standard problems in a new context or adding an extra layer of complexity to the scenario.
- Emphasis on Deep Understanding: Questions should assess not only procedural skill but also the implications and reasoning behind the concepts.

2. Content Coverage and Difficulty Distribution:
- Cover all major topics from the course (course name, description, syllabus, and/or exam).
- In case of any mismatch between the course description/syllabus and exam content, default to using the course name and description.
- Ensure each question assesses a unique concept.
- The overall set should be challenging with an average student scoring around 50%.
- Difficulty distrubution. Generate questions that would be seen as medium difficulty in an AP exam of for the given course. Example questions for Statistics are provided below.

Calibration Examples from a Statistics Course:
{AP STATS QUESTIONS, OMITTED}

PAST JUDGE DECISIONS:

GOOD EXAMPLE #{number}:
QUESTION: {question}
OPTIONS: {answer_options}
DECISION: KEEP

BAD EXAMPLE #{number}:
QUESTION: {question}
OPTIONS: {answer_options}
DECISION: REMOVE

\end{lstlisting}
\begin{lstlisting}[
  basicstyle=\ttfamily,
  numbers=none,
  frame=lines, 
  xleftmargin=2pt,  
  breaklines=true,
  breakautoindent=false,
  breakindent=0pt,
  breakatwhitespace=true,
  title={Self-Refine Judge Prompt}
]
You are judging proposed exam questions for a course. You have the following course info:

Course Name: {course_name}
Course Description: {desc}
Exam Content: {exam_content}
Syllabus Content: {syllabus_content}

{course_info}

Below are some previously judged examples:

GOOD EXAMPLE #{number}:
QUESTION: {question}
OPTIONS: {answer_options}
DECISION: KEEP

BAD EXAMPLE #{number}:
QUESTION: {question}
OPTIONS: {answer_options}
DECISION: REMOVE

Here is the proposed question:

Q: {question}
Options: {options}

Appropriateness guidelines:

- No direct references to explicit syllabus content. (e.g. "Which statistical topic is NOT explicitly emphasized in the STAT 1: Introduction to Statistics course based on the syllabus description?")
- No questions about course logistics (term paper requirements, office hours).
- No references to external resources or tables (z-tables, software, etc.).
- Any calculations must be solvable mentally (no calculator, pen and paper, or tables).
- Any question that is purely formula-plugging or trivial is inappropriate.
- Answer options must not refer to one another (e.g. "Option A is correct and B is False").
- The question must relate to core concepts reasonably inferred from the course description, syllabus, and/or exams.
- You should be very strict about this. If a concept does not appear in the provided materials, we should never have an exam question about it.

Appropriateness example:

**Question:** The number of misprints on a page of a newspaper follows a Poisson distribution 
with a mean of 1.2 errors per 1000 words. What is the probability of finding exactly 4 errors in 1000 words?

This is inappropriate because it only asks students to apply a memorized formula.
   
Calibration Examples from a Statistics Course: {Omitted}

Other rejection criteria:
- If the question asks about a concept that is already covered in the GOOD examples (above in the few-shot block), remove it.
- Questions should NOT ask about concepts that have already been covered. 
- If it rehashes or duplicates the BAD examples, remove it.

Decide if we KEEP or REMOVE the question:
- Return exactly "Keep" if the question meets guidelines.
- Otherwise return "Remove".
\end{lstlisting}
\subsection{Example Exam: Programming Languages}
\begin{lstlisting}[
  basicstyle=\ttfamily,
  numbers=none,
  frame=lines, 
  xleftmargin=2pt,  
  breaklines=true,
  breakautoindent=false,
  breakindent=0pt,
  breakatwhitespace=true,
  title={Programming Languages Exam}
]
Question 1/10
Which of the following is NOT a typical benefit of using a virtual machine (VM) when implementing programming languages?
a. Platform independence for running code on different hardware
b. Enhanced runtime optimizations through techniques like JIT compilation
c. Direct access to low-level hardware registers and machine-specific instructions
d. Improved security and isolation of the execution environment

Question 2/10
In designing a programming language that integrates multiple paradigms (imperative, object-oriented, functional, and declarative), which of the following strategies best facilitates a consistent semantic specification while enabling efficient execution on a virtual machine?
a. Using separate compilers for each paradigm
b. Leveraging a common abstract machine with unified control structures
c. Restricting the language to a single paradigm to simplify semantics
d. Relying solely on syntax-directed translation for code generation

Question 3/10
In ML, the functions foldl and foldr are used for list processing. Which of the following statements best summarizes the trade-offs between foldl and foldr in a strict evaluation language like ML?
a. foldl is tail-recursive and more memory-efficient for accumulating results on large lists while foldr is generally non-tail-recursive and may cause stack overflow on very long lists.
b. foldr is tail-recursive and thus more efficient for processing lists whereas foldl is non-tail-recursive and prone to performance issues on large lists.
c. Both foldl and foldr are tail-recursive in ML so the choice between them only affects the order of element processing not performance.
d. Neither foldl nor foldr is tail-recursive in ML making both functions equally inefficient for processing very long lists.

Question 4/10
In a language design process that employs separate syntactic and semantic specifications, which potential issue is most likely to occur if updates to the formal grammar (e.g., using BNF) are not properly reflected in the corresponding semantic rules?
a. The compiler may become more efficient due to fewer semantic checks.
b. Ambiguities in the grammar will be resolved automatically during parsing.
c. The runtime system will ignore outdated semantic actions and use default behaviors.
d. Inconsistent behavior may result as the refactored grammar no longer aligns with the semantic actions.

Question 5/10
In a statically-typed functional language that is extended with object-oriented features, what is a primary design challenge that arises from such integration?
a. Ensuring functions remain pure despite the presence of mutable objects
b. Eliminating the need for compile-time type checking
c. Simplifying method overloading to enhance runtime performance
d. Reducing the overall size of the runtime system by merging paradigms

Question 6/10
In designing a virtual machine for a multi-paradigm programming language that supports both functional and object-oriented styles, which design decision best enables the efficient execution of lazy evaluation and dynamic dispatch?
a. Implementing separate runtime systems for functional and object-oriented code
b. Using a unified runtime environment that supports deferred computations along with method tables for dynamic dispatch
c. Prioritizing eager evaluation for all constructs to simplify execution
d. Relying solely on just-in-time compilation to handle both paradigms

Question 7/10
In the design of a programming language, what is a primary benefit of clearly separating its syntactic specification (for example, using a formal grammar like BNF) from its semantic specification (such as type rules and execution behavior)?
a. It allows independent evolution and modification of syntax and semantics without impacting each other
b. It enables direct code execution without any intermediate translation steps
c. It reduces the need for runtime error checking by merging syntax and semantic analysis
d. It simplifies parsing by eliminating the need for a separate lexical analyzer

Question 8/10
When specifying a programming language's syntax with Backus-Naur Form (BNF), which technique is most effective in reducing ambiguity and simplifying parsing?
a. Increasing the number of production rules to cover diverse constructs
b. Implementing left-recursive productions to enforce recursion
c. Applying left-factoring to isolate common prefixes in production rules
d. Using separate grammars for distinct paradigms and merging them at parse time

Question 9/10
Which of the following best describes the type inference mechanism used in ML?
a. It relies on runtime type checking to determine the type of each expression
b. It automatically deduces the most general type for expressions without explicit type annotations
c. It requires every function and variable to be explicitly declared with their types
d. It applies only to primitive types leaving user-defined types unchecked

Question 10/10
In designing a multi-paradigm programming language that integrates imperative, object-oriented, functional, and declarative styles, which design aspect requires the most careful consideration to ensure consistent behavior across all paradigms?
a. Memory allocation strategies in the runtime system
b. Variable scoping and binding rules
c. Syntax for loop and control structures
d. Low-level hardware interfacing capabilities
\end{lstlisting}

\end{document}